\input harvmac.tex



\let\includefigures=\iftrue
\newfam\black
\includefigures
\input epsf
\def\figin{\epsfcheck\figin}\def\figins{\epsfcheck\figins}
\def\epsfcheck{\ifx\epsfbox\UnDeFiNeD
\message{(NO epsf.tex, FIGURES WILL BE IGNORED)}
\gdef\figin##1{\vskip2in}\gdef\figins##1{\hskip.5in}
\else\message{(FIGURES WILL BE INCLUDED)}%
\gdef\figin##1{##1}\gdef\figins##1{##1}\fi}
\def\DefWarn#1{}
\def\figinsert{\goodbreak\midinsert}
\def\ifig#1#2#3{\DefWarn#1\xdef#1{fig.~\the\figno}
\writedef{#1\leftbracket fig.\noexpand~\the\figno}%
\figinsert\figin{\centerline{#3}}\medskip\centerline{\vbox{\baselineskip12pt \advance\hsize by
-1truein\noindent\footnotefont{\bf Fig.~\the\figno:} #2}}
\bigskip\endinsert\global\advance\figno by1}
\else
\def\ifig#1#2#3{\xdef#1{fig.~\the\figno}
\writedef{#1\leftbracket fig.\noexpand~\the\figno}%
\global\advance\figno by1} \fi

\font\cmss=cmss10 \font\cmsss=cmss10 at 7pt

\def\IB{\relax\hbox{$\inbar\kern-.3em{\rm B}$}}
\def\IC{\relax\hbox{$\inbar\kern-.3em{\rm C}$}}
\def\IQ{\relax\hbox{$\inbar\kern-.3em{\rm Q}$}}
\def\ID{\relax\hbox{$\inbar\kern-.3em{\rm D}$}}
\def\IE{\relax\hbox{$\inbar\kern-.3em{\rm E}$}}
\def\IF{\relax\hbox{$\inbar\kern-.3em{\rm F}$}}
\def\IG{\relax\hbox{$\inbar\kern-.3em{\rm G}$}}
\def\IGa{\relax\hbox{${\rm I}\kern-.18em\Gamma$}}
\def\IH{\relax{\rm I\kern-.18em H}}
\def\IK{\relax{\rm I\kern-.18em K}}
\def\IL{\relax{\rm I\kern-.18em L}}
\def\IP{\relax{\rm I\kern-.18em P}}
\def\IR{\relax{\rm I\kern-.18em R}}
\def\Z{\relax\ifmmode\mathchoice
{\hbox{\cmss Z\kern-.4em Z}}{\hbox{\cmss Z\kern-.4em Z}} {\lower.9pt\hbox{\cmsss Z\kern-.4em Z}}
{\lower1.2pt\hbox{\cmsss Z\kern-.4em Z}}\else{\cmss Z\kern-.4em Z}\fi}

\def\II{\relax{\rm I\kern-.18em I}}

\def\S{{\bf S}}

\def\cp{{\bf CP}}

\def\AdS{{\rm AdS}}


\def\CF {{\cal F}}

\def\CO {{\cal O}}


\def\p{\partial}

\def\tilde{\widetilde}

\def\bar{\overline}


\def\p{\partial}

\def\inbar{\,\vrule height1.5ex width.4pt depth0pt}
\def\r{{\rm Re}}
\def\i{{\rm Im}}

\def\a{\alpha}
\def\b{\beta}

\def\d{\delta}

\def\th{\theta}

\def\bar{\overline}

\def\det{{\rm det}}

\def\IH{{\bf H}}


\lref\Douglas{
  M.~R.~Douglas,
  ``Basic results in vacuum statistics,''
  Comptes Rendus Physique {\bf 5}, 965 (2004)
  [arXiv:hep-th/0409207].
  }

\lref\KKLT{
  S.~Kachru, R.~Kallosh, A.~Linde and S.~P.~Trivedi,
  ``De Sitter vacua in string theory,''
  Phys.\ Rev.\ D {\bf 68}, 046005 (2003)
  [arXiv:hep-th/0301240].
}

\lref\HH{
  J.~B.~Hartle and S.~W.~Hawking,
  ``Wave Function Of The Universe,''
  Phys.\ Rev.\ D {\bf 28}, 2960 (1983).}

\lref\ovv{H.~Ooguri, C.~Vafa and E.~Verlinde, ``Hartle-Hawking wave-function for flux compactifications,''
hep-th/0502211.}

\lref\osv{H.~Ooguri, A.~Strominger and C.~Vafa, ``Black hole attractors and the topological string,'' Phys.\ Rev.\ D
{\bf 70}, 106007 (2004) [arXiv:hep-th/0405146].}

\lref\cano{ P.~Candelas and X.~de la Ossa,
  ``Moduli Space Of Calabi-Yau Manifolds,''
  Nucl.\ Phys.\ B {\bf 355}, 455 (1991).}

\lref\attra{ S.~Ferrara, R.~Kallosh and A.~Strominger,
  ``N=2 extremal black holes,''
  Phys.\ Rev.\ D {\bf 52}, 5412 (1995).
}

\lref\GSVY{B. Greene, A. Shapere, C. Vafa, and S.-T. Yau, ``Stringy Cosmic Strings and Noncompact Calabi-Yau
Manifolds,'' Nucl. Phys. {\bf B337} (1990) 1--36.}

\lref\COGP{P.~Candelas, X.~C.~De La Ossa, P.~S.~Green and L.~Parkes,
  ``A Pair Of Calabi-Yau Manifolds As An Exactly Soluble Superconformal
  Theory,''
  Nucl.\ Phys.\ B {\bf 359}, 21 (1991).
  }

\lref\Ferrara{E.~Cremmer, C.~Kounnas, A.~Van Proeyen, J.~P.~Derendinger, S.~Ferrara, B.~de Wit and L.~Girardello,
  ``Vector Multiplets Coupled To N=2 Supergravity: Superhiggs Effect, Flat
  Potentials And Geometric Structure,''
  Nucl.\ Phys.\ B {\bf 250}, 385 (1985).
  }

\lref\GoV{
  R.~Gopakumar and C.~Vafa,
  ``M-theory and topological strings. I, II''
  arXiv:hep-th/9809187,
  arXiv:hep-th/9812127.
}

\lref\asp{P.~S.~Aspinwall, ``Compactification, geometry and duality: N = 2,'' arXiv:hep-th/0001001.}

\lref\KZ{A.~Klemm, E.~Zaslow, ``Local Mirror Symmetry at Higher Genus,'' hep-th/9906046.}

\lref\DG{D.~E.~Diaconescu and J.~Gomis, ``Fractional branes and boundary states in orbifold theories,'' JHEP {\bf
0010}, 001 (2000), hep-th/9906242.}

\lref\AKV{M.~Aganagic, A.~Klemm, C.~Vafa, ``Disk Instantons, Mirror Symmetry and the Duality Web,'' hep-th/0105045.}

\lref\Mumford{D.~Mumford, ``A note on Shimura's paper {\it Discontinuous Groups and Abelian Varieties},'' Math. Ann.
{\bf 181} (1969) 345.}

\lref\Borcea{C.~Borcea, ``Calabi-Yau Threefolds and Complex Multiplication,'' in {\it Essays on Mirror Manifolds,}
S.-T.~Yau ed., International Press, 1992.}

\lref\Moore{G.~Moore, ``Arithmetic and attractors,'' arXiv:hep-th/9807087.}

\lref\GV{S.~Gukov and C.~Vafa, ``Rational conformal field theories and complex multiplication,'' Commun.\ Math.\ Phys.\
{\bf 246}, 181 (2004), hep-th/0203213.}

\lref\AMV{M.~Aganagic, M.~Marino and C.~Vafa, ``All loop topological string amplitudes from Chern-Simons theory,''
Commun.\ Math.\ Phys.\  {\bf 247} (2004) 467, hep-th/0206164.}

\lref\SYZ{A.~Strominger, S.~T.~Yau, and E.~Zaslow, ``Mirror Symmetry is T-duality,'' Nucl. Phys. {\bf B479} (1996) 243,
hep-th/9606040.}

\lref\bateman{The Bateman Project, {\sl Higher Transcendental Functions,} Vol. 1, Sec. 5.3.-5.6, A. Erdelyi ed.,
McGraw-Hill Book Company, New York (1953).}

\lref\dgr{F.~Denef, B.~R.~Greene and M.~Raugas,
  ``Split attractor flows and the spectrum of BPS D-branes on the quintic,''
  JHEP {\bf 0105}, 012 (2001)
}

\lref\GSV{S.~Gukov, K.~Saraikin and C.~Vafa,
``A stringy wave function for an $S^3$ cosmology,'' hep-th/0505204.}

\lref\Riema{B.~Wajnryb, ``Mapping class group of a surface is generated by two elements,''
Topology {\bf 35} (1996) 377.}

\lref\Riemb{M.~Korkmaz, Trans. Amer. Math. Soc. {\bf 357} (2005) 3299.}


\Title{\vbox{\baselineskip11pt\hbox{hep-th/0509109}
\hbox{HUTP-05/A041}
\hbox{ITEP-TH-55/05}
}}
{\vbox{
\centerline{The Entropic Principle and Asymptotic Freedom}
}}
\centerline{
Sergei Gukov$^{1,2}$\footnote{$^{\dagger}$}{On leave from: ITEP,
Moscow, 117259, Russia and
L.D.Landau ITP, Moscow, 119334, Russia},
Kirill Saraikin$^{2\dagger}$,
and Cumrun Vafa$^2$}
\medskip
\medskip
\medskip
\vskip 8pt
\centerline{$^1$ \it California Institute of Technology 452-48,
Pasadena, CA 91125, USA}
\medskip
\centerline{$^2$ \it
Jefferson Physical Laboratory, Harvard University, Cambridge, MA
02138, USA}
\medskip
\medskip
\medskip
\noindent
Motivated by the recent developments about the Hartle-Hawking
wave function associated to black holes, we formulate an entropy
functional on the moduli space of Calabi-Yau compactifications.
We find that the maximization of the entropy
is correlated with the appearance of asymptotic freedom in the effective
field theory.
The points where the entropy is maximized correspond to points
on the moduli which are maximal intersection points of walls of marginal
stability for BPS states.  We also find an intriguing link between
extremizing the entropy functional and the points on the moduli
space of Calabi-Yau three-folds which admit a `quantum deformed'
complex multiplication.

\medskip
\Date{September 2005}

\newsec{Introduction}

There is little doubt that there exists a large number of consistent
superstring vacua.  This fact is not new; it has been well
known for a while in the context of supersymmetric vacua.
More recently, there has been some evidence
that the multitude of vacua continues to exist
even without supersymmetry
(for introduction and references see \refs{\KKLT, \Douglas}).
Of course, one can stop here and resort to the standard
philosophy of physics:  Choose the theory to be in accord
with observation.  However, in the context of string theory,
being a unified theory of all matter, it is natural to explore
whether one can say a little more about the selection criteria.
The purpose of this paper is to further explore one idea
along these lines, advanced recently \ovv.

The idea in \ovv\ is to interpret the results of \osv\
in the context of flux compactifications on $\AdS_2\times \S^2 \times M$,
where $M$ is a Calabi-Yau threefold (and where for simplicity we ignore
a $\Z$-identification).
The norm of the Hartle-Hawking wave function associated with
this background can be interpreted holographically as the black hole entropy.
In particular, the flux data on the $\AdS_2\times \S^2$ geometry is
mapped to the charge of the dual black hole, and the norm of the
wave function satisfies
\eqn\psipsis{ \langle \psi | \psi \rangle ={\rm exp}(S) }
where $S$ denotes the entropy of the corresponding black hole.
For a fixed flux data, the wave function $\psi$ can be
viewed as a function over the moduli space of the Calabi-Yau,
together with the choice of the normalization for the holomorphic
3-form, where the overall rescaling
of the holomorphic 3-form corresponds to the overall rescaling
of the charge of the black hole.  Clearly the entropy
of the black hole increases as we rescale the overall charge.
However, to obtain a wave function
on the Calabi-Yau moduli space, one would like to get rid of this extra
rescaling.  The main purpose of this paper is to suggest one mechanism
of how this may be done:  We simply fix one of the magnetic charges, and its
electric dual chemical potential.
 In this way, as we shall
argue, the wave function becomes a function on the geometric moduli space
of the Calabi-Yau and one can see which Calabi-Yau manifolds are ``preferred''.

It turns out that this problem can be formulated for both compact and non-compact
Calabi-Yau manifolds. It is a bit more motivated
in the non-compact case, since in this case
there is a canonical choice of the fixed charge
(D0 brane in the type IIA context).  We find that the condition
for a maximum/minimum corresponds to the points of intersection
of walls of marginal stability.  Moreover we find  a set of solutions
in various examples.
One type of solutions we find corresponds
to points on moduli space which admit a complex multiplication type structure.
 We also find examples where extrema correspond to
the appearance of extra massless particles. We find that in these
cases the norm of the wave function $\psi$ is {\it maximized} in
correlation with the sign of beta function: Asymptotically free
theories yield maximum norm for the wave function.

The organization of this paper is as follows: we start  in section 2
with the formulation of the problem. Then, in section 3,
we find the conditions
for maxima/minima of the wave function.  We also explain
why this favors asymptotically free theories in local examples where
there are massless fields.
In section 4 we give examples of our results and
in section 5 we end with conclusions and some open questions.

\newsec{General Formulation of the Problem}

Consider compactifications of type IIB superstrings on $\AdS_2 \times \S^2 \times M$
where $M$ is a Calabi-Yau threefold, which may or may not be compact.  In addition
we consider fluxes of the 4-form gauge field along $\AdS_2 \times \S^2$
and 3-cycles of $M$.  Let $F_{p,q}=p^I\alpha_I+q_J \beta^J$ denote the flux
through $M$, where
$\alpha_I$ and $\beta^J$ form a canonical symplectic basis for integral
3-form cohomology $H^3(M,\Z)$.  According to \ovv , the results
of \osv\ can be interpreted in terms of a Hartle-Hawking type wave function $\psi_{p,q}$
for this geometry on the minisuperspace,
with the property that
\eqn\psipqs{ \langle \psi_{p,q}|\psi_{p,q}\rangle = \exp S(p,q) }
Here
$S(p,q)$ denotes the entropy of the dual black hole obtained
by wrapping a D3 brane with magnetic and electric charges
$p$ and $q$.

In the limit of large fluxes,
\eqn\pqscaling{ (p,q)\rightarrow \lambda (p,q) }
where $\lambda >>1$, the entropy $S(p,q)$ has a classical
approximation, given by the Bekenstein-Hawking formula.  In this
limit, the curvature of $\AdS_2 \times \S^2$ becomes small,
and the entropy of the black hole is equal to ${1 \over 4} A(\S^2)$.
In particular, the attractor mechanism \attra\ will
freeze the complex structure moduli of the Calabi-Yau space $M$,
so that there exists a holomorphic 3-form $\Omega$ on $M$
with the property
\eqn\fixo{ \r (\Omega)=F_{p,q}}
Moreover, in this limit the entropy is given by
$$S(p,q)={A(\S^2)\over 4}=-i {\pi \over 4}\int_M \Omega \wedge {\overline \Omega}$$
where $\Omega$ fixed by \fixo .  Furthermore, in this limit, $ \i \Omega$
plays the role of the chemical potential.

Suppose we wish to ask the following question: In type IIB compactification
on $\IR^4\times M$, which Calabi-Yau $M$ is ``preferred''?
One way to tackle this question is to embed it in the geometry
$\AdS_2 \times \S^2\times M$ where the complex structure
of $M$ is determined by the fluxes, through the attractor mechanism. Then, the question becomes:
For which values of the complex structure moduli
the norm of the wave function is maximized or, in other words,
for which attractor Calabi-Yau the entropy
of the corresponding black hole is maximized?
In this formulation of the question,
we can view $\IR^4$ as a special limit of $\AdS_2\times \S^2$ where
the charge of the black hole is rescaled by an infinite amount
$\lambda \rightarrow \infty$.
Thus $\AdS_2\times \S^2$ can be viewed as a regulator geometry for $\IR^4$.

However, this way of asking the question leads to the following pathology:
The entropy of the black hole for large $\lambda $ scales
as $\lambda^2$.  Therefore, in order to get a reasonable
function on the moduli space of $M$ we need to fix the normalization of $\Omega$.
One way to do this is to fix the value of $\int \Omega \wedge {\overline \Omega}$
so that it is the same at all points in moduli; however this is precisely the entropy that we wish to maximize.
  If we fixed the normalization of $\Omega$ in this way, we would obtain, tautologically, a flat distribution on the
moduli space of $M$.  In this
sense there would be no particular preference of one point on the moduli of $M$
over any other.
Instead we consider the following mathematically natural alternative:
We choose a 3-cycle $A_0\subset M$ and fix the normalization
of $\Omega$ by requiring it to have a fixed period along $A_0$,
\eqn\fixi{\int_{A_0} \Omega ={\rm fixed}}
Since the overall scale of $\Omega $ does not affect the extremum point on
the moduli space of $M$, with no loss of generality we can fix the above period to be $1$.
Then, we can  consider maximizing the entropy, which now depends only
on the geometric moduli of $M$. Thus our problem becomes
\eqn\prob{{\rm Maximize} \ \Big|\int \Omega \wedge {\overline \Omega}\ \Big|\qquad
{\rm subject}\
{\rm to}\ ~~~  \int_{A_0} \Omega =1.
}
Physically, what this means is that we fix one of the charges of the black hole,
say the magnetic charge $p^0=1$, and the corresponding electric chemical potential
$\phi^0=0$ (determined by the imaginary part of $\Omega$).

Strictly speaking, the above problem is well defined on the moduli space
of Calabi-Yau manifolds together with a choice of a 3-cycle.
This, in general, is a covering of the moduli space.  Nevertheless,
any particular maximization of the entropy functional on this covering
space will descend to a particular choice of the complex structure moduli of $M$
(by forgetting on which sheet the function is maximized).
Of course, it would be interesting to find out whether or not this covering
of the moduli space is a finite covering or not.
We will discuss some aspects of this in section 5.
We should point out, however, that there is a {\it canonical}
choice of the cycle $A_0$ in the mirror type IIA problem,
where one is studying even-dimensional D-branes wrapped over
even-dimensional cycles of a non-compact Calabi-Yau manifold.
In this case, one can choose $A_0$ to be a point on $M$ and
consider a fixed number of $D6$ branes with zero
chemical potential for D0 branes.
We also note that, in the type IIA setup, fixing one of the periods can be interpreted
as fixing the topological string coupling constant
\eqn\xnot{ X^0 = {4 \pi i\over g_s} }
In what follows, we consider both compact and non-compact examples.

\newsec{General Conditions for Maxima/Minima}

In this section, we derive equations for the critical points of
the constrained variational problem described in
the previous section, and discuss maxima of the entropy functional
\eqn\entropy{\eqalign{ S = -i {\pi \over 4} \int_{M} \Omega \wedge \bar \Omega
 }}
It turns out that the critical points of \entropy\ are described by
equations of the form
$$\i a^D - \tau \, \i a= 0, $$
 where $(a, a^D)$ denote
``reduced'' periods of the Calabi-Yau and $\tau = {da^D \over da}$ is the coupling constant
matrix. The points on the moduli space where the entropy is
maximized are those where $\i \tau >0$ and all but one of the
Calabi-Yau periods have equal phase. As we explain below,
these are also the points where the maximal number of
the walls of marginal stability for BPS states meet together.
Notice, the restriction $\i \tau >0 $ implies that
the effective field theory for the extra massless particles appearing
at the maximum point can be decoupled from the gravity.
This is a necessary condition for the
asymptotically free effective theories.

\subsec{Critical Points}

In order to extremize the functional \entropy\ we
need to introduce coordinates on the
 moduli space of a Calabi-Yau manifold with one fixed 3-cycle.
Given the symplectic basis of 3-cycles
$\{A_I, B^J\}_{I,J=0, \ldots, h^{2,1}}$,
such that $\# (A_I, B^J) = \delta_I^{\ J}$,
the periods of the holomorphic 3-form are
\eqn\aper{\eqalign{
X^I = \int_{A_I} \Omega
}}
\eqn\bper{
\eqalign{ F_I = \int_{B^I} \Omega
}}
In particular, one can use $X^I$ as the  projective coordinates
on the Calabi-Yau moduli space, and express the $B$-periods
as derivatives of the prepotential:
\eqn\prepx{
F_I(X) = {\p \CF_0(X) \over \p X^I}
}
We choose the fixed 3-cycle to be $A_0$.
Then the normalization of $\Omega$ is fixed by the condition
\eqn\xnt{ X^0 = \int_{A_0} \Omega }
As we discussed earlier, we can always set $X^0=1$.
However, in what follows it will be useful to keep the dependence on $X^0$
which, in the type IIA context, determines the topological string coupling constant,
{\it cf.}  \xnot.

It is natural to use the following  coordinates
on the moduli space of Calabi-Yau manifolds with a fixed 3-cycle:
\eqn\ai{ a^i = {X^i \over X^0}, \qquad  i = 1, \ldots, h^{2,1} }
Also, we introduce the ``dual'' variables
\eqn\aid{ a_i^D =  {F_i \over X^0} }
and a ``rigid" prepotential
\eqn\prepa{ F(a) = \big(X^0\big)^{-2} \CF_0\big(X(a)\big) }
Then, using the fact that $\CF_0$ is a homogeneous holomorphic
function of degree two, we find
\eqn\fnot{ F_0 = X^0 ( 2F - a^i a^D_i) }
Therefore,
the functional \entropy\
can be written as\foot{
Here we used the Riemann bilinear identity
$
\int_{M} \a \wedge \b = \sum_I
\big( \int_{A_I} \a \int_{B^I} \b - \int_{A_I} \b \int_{B^I} \a \big)
$.}
\eqn\entropya{\eqalign{ S = i {\pi \over 4} \big|X^0 \big|^2 \big\{ 2(F-\bar F) -
(a^i - \bar a^i) (a_i^D + \bar a_i^D)
\big\} }}
This is, of course, the standard
expression for the K\"{a}hler potential $S={\pi \over 4}\, e^{-K}$, written in
terms of the special
coordinates, see {\it e.g.} \cano.

Extremizing the action \entropya\ with respect to $ a^i$ and $\bar a^i$,
we obtain the following system of equations:
\eqn\cm{ \i a_i^D - \tau_{ij} \i a^j = 0 }
where the coupling constant matrix $\tau_{ij}$ is given by
\eqn\taudef{ \tau_{ij} = {\p^2 F \over \p a^i \p a^j }
\equiv {\p^2 \CF_0 \over \p X^i \p X^j } }
Solutions to the  equations \cm\ define critical points on the Calabi-Yau moduli space.
Our goal will be to
study these points and to understand their physical and/or geometric meaning.

Before we proceed, let us make a few general comments about the form
of the equations \cm.
First, note that \cm\ is a system of non-linear complex equations.
Even though these equations are not differential, finding their
solutions is a challenging and interesting problem.
To see this, let us write \cm\ in the following form
\eqn\cmviaf{ \i \, \p_i F -  (\p_i \p_j F) \i a^j = 0 }
where we expressed $a^D_i$ and $\tau_{ij}$ in
terms of the single function $F(a^i)$. For a given Calabi-Yau space $M$
and a choice of the 3-cycle, the function $F (a^i)$ is fixed;
it is generically a non-trivial transcendental function. Therefore, \cm\
(equivalently \cmviaf)
represents a system of $n=h^{2,1}$ complex equations for $n$ complex
variables $a^i$, $i=1,\ldots,n$. Therefore one expects that solutions to
these equations are isolated points in the moduli space.

Let us note that Calabi-Yau manifolds which correspond
to these points
admit special structures, analogous to the complex multiplication.
The notion of complex multiplication for higher dimensional
varieties goes back to the work of Mumford \Mumford;
in the context of Calabi-Yau three-folds it was studied by Borcea \Borcea.
In the physics literature, it appears in the study of black hole
attractors \Moore\ and rational conformal field theories \GV.
Let us recall the attractor equations \attra\
\eqn\attr{\eqalign{
2 \i a^i & = p^i \cr
2 \i a_i^D & = q_i
}}
where $(p,q) \in \Z^n$ denote magnetic and electric fluxes.
One says that a Calabi-Yau manifold admits complex multiplication
if the Jacobian $ T = \IC^n /(\Z^n + \tau \Z^n)$ associated
with the coupling constant matrix
$\tau_{ij}$ admits complex multiplication. This occurs if
$\tau_{ij}$ satisfies
the following second order matrix equation:
\eqn\quadeq{
\tau C \tau + A\tau -\tau D - B = 0
}
where $A,B,C,D$ are some integer matrices.
It is straightforward to check that this is indeed  the case
for a suitable choice of the integer matrices\foot{For example,
$
A = n \vec q \otimes  \vec q,
\qquad
B = 0,
\qquad
C = m \vec p \otimes  \vec p - n \vec p \otimes  \vec q,
\qquad
D = m \vec p \otimes  \vec q
$
}.
Therefore, any Calabi-Yau with moduli fixed by the attractor
mechanism \attr\ and satisfying equation \cm\
admits complex multiplication.
Notice, that Jacobian $T$ for this Calabi-Yau is singular,
since \cm\ implies that there are fixed points under
the $\Z^n + \tau \Z^n$ action.

Solutions to \cm\ fall into three families, which can be characterized
by the imaginary part of the coupling constant
matrix $\tau_{ij}$. To see this, notice that the
imaginary part of extremum equations \cm\ is given by
\eqn\imextr{ \i \tau_{ij} \cdot \i a^j = 0 }
Therefore, if $\i \tau_{ij}$ is non-degenerate
(that is if $\det || \i \tau_{ij} || \not = 0$), the only possible solution
is $\i a^i = 0$. Moreover, assuming that $\tau_{ij}$ remains
finite\foot{or if $\tau \sim a^{-\a}$, where $\a<1$}, it also
follows that $\i a^D_i = 0$. We shall refer to this family of solutions
as solutions of type I:
\eqn\triv{ \i a^i = 0, \quad \quad \i a^D_i = 0,\quad \quad \   i = 1, \dots, n }
The expression  for the entropy functional calculated
at the critical point of type I  turns out to be very simple:
\eqn\entrcr{ S_*= i {\pi \over 2} \big|X^0 \big|^2 (F-\bar F) }
If we go to the conventional topological string notations
\eqn\fif{ F_{top}=i(2\pi)^3 F, \ \ Z = \exp {1 \over g_s^2}F_{top} }
and use \xnot, we see that the probability function for the Calabi-Yau
at the critical point is given by the
square of the topological wave function $\Psi_{top} = Z$,
in accordance with \refs{\ovv,\osv}:
\eqn\probb{ e^{S_*} = |\Psi_{top}|^2 }
Solutions of type II correspond to $\det || \i \tau_{ij} ||  = 0$.
They can be expressed in terms of
{\it real} eigenvectors $v^i$ of the coupling constant matrix,
$\i \tau_{ij} \cdot v^j = 0$,
\eqn\ntriv{ \i a^i = v^i, \quad \quad \i a^D_i = \tau_{ij} v^i ,
\quad \quad \   i,j = 1, \dots, n }
Finally, solutions of type III correspond to divergent coupling constant matrix:
\eqn\div{ \i a^i = 0, \quad \quad \i a^D_i =\lim_{ \i a^i \to 0}
\r \tau_{ij} \cdot \i a^i
\not = 0,\quad \quad \   i = 1, \dots, n }
All types of the solutions represent some special points on the
Calabi-Yau moduli space.

\subsec{Conditions for Maximum/Minimum}

It is natural to ask which of these
critical points are maxima and which are minima. The former
correspond to theories which are preferred, while the latter
correspond to the theories which are least likely, according
to the entropic principle. Therefore,
our main goal is to search for the maximum points.

In order to answer this question, we need to look at
the second variation of the action at the critical point:
\eqn\deltas{ \d^{2} S = - {\pi \over 2} \big|X^0
\big|^2 ( \i \tau_{ij} ) \d a^i \d \bar a^j + {\pi \over 4} \big|X^0
\big|^2 \i a^i \big( c_{ijk} \d a^j \d a^k + \bar c_{ijk} \d \bar
a^j \d \bar a^k\big) }
where $c_{ijk}$  are defined as
\eqn\taudef{ c_{ijk} = {\p^3 F \over \p a^i \p a^j \p a^k} }
The bilinear form $\i a^i c_{ijk} \d a^j \d a^k$
does not have a definite signature. This means
that if it is non-zero, the critical point is neither
minimum nor maximum.
Therefore, a necessary condition for the critical point
to be a local maximum is
\eqn\mnmx{
 c_{ijk}  \i a^k =0
}
Let us concentrate  on the first term in \deltas,
assuming that  this condition is satisfied.
Remember that reduced coupling constant matrix
$\tau_{ij}$ is part of a full matrix $\tau_{IJ}$.
Imaginary part of this matrix has signature\foot{
The simplest illustration is a toy model with the
cubic prepotential, $\CF_0 =- {1\over 3} {( X^1)^3 \over X^0} $.
It is easy to check that the signature of $\i \tau_{IJ}$
in this example is (1,1). Notice, that there is a difference
between the physical coupling constant matrix, including
the graviphoton coupling,
 and $\tau_{IJ}$. The physical coupling constant matrix  is always positive definite
 (see \Ferrara).
}
of type $(h^{2,1},1)$, as follows from the expression
\eqn\itau{
\i \tau_{IJ}= {i\over 2}\int \p_I \Omega \wedge \bar \p_J \bar \Omega
}
and  decomposition $\p_I \Omega \in H^{2,1}(M) \oplus H^{3,0}(M)$.
Therefore, the signature of the reduced  matrix $\tau_{ij}$ is either $(h^{2,1},0)$
or $(h^{2,1}-1,1)$. In the first case the form
$(\i \tau_{ij}) \d a^i \d \bar a^j $ is positive definite
and therefore, the entropy functional has
\eqn\max{\eqalign{ { maximum}, \ \ {\rm if} \ \  \i \tau_{ij}>0
 }}
In the second case  generically we have
a saddle point, and a minimum if $h^{2,1}=1$.
Thus we conclude that, in general, Calabi-Yau models with
$\i \tau_{ij}>0$ are preferred.

The extremum conditions  \cm\  are very restrictive,
but it is hard to find a solution in general case.
However,  if we want to satisfy constraint \mnmx,
which is necessary for maximization of the entropy, the problem
simplifies. Assuming that $c_{ijk}$ is not of special degenerate form,
a general solution
to this constraint is $\i a^i = 0$, and therefore
we should look at the type I solution
\eqn\trv{
\i a^i = \i a^D_i = 0
}
Physically, this is a particularly natural choice and we can explain it
in yet another way:  The extremum of the probability density given
by the entropy functional, should naturally pick attractor fixed points.
In our problem the electric chemical potential is set to zero and the
magnetic charge is fixed in one particular direction.  It is not surprising
that this means that the rest of the charges are set to zero at the extremum.
In particular the equation \trv\ can naturally be interpreted
as the attractor point with this set of charges.

 Another way to derive \trv\
is to notice that imaginary part of the coupling constant matrix
enters the   second variation of the action
 \deltas\ and therefore it should be non-degenerate
at the local minimum or maximum point:
\eqn\imtau{
\det || \i \tau_{ij} || \not = 0
}
Combining this with the imaginary part of extremum equations  \imextr,
we get $\i a^i = 0$.

One of the interesting examples of the maximum entropy solutions
is the one with the logarithmic behavior of the
coupling constant matrix (explicit examples of this
are discussed in the next section):
\eqn\taubeta{
\tau = \tau_0 + i \beta \log {a^2\over \Lambda^2} }
near the critical point $a=0$. From the point
of view of the corresponding effective field theory in four dimensions
this expression describes an  RG flow of the couplings, and $\beta$ is the one-loop beta
function of the effective field theory near the point $a=0$.
Combining this with the maximum
condition $\i \tau_{ij}>0$ discussed earlier, we conclude that
for the theories with $\beta < 0$, that is, {\it for asymptotically free theories
the probability density is maximized}.

\subsec{Marginal Stability Curves and the Entropy}

As we will  explain below, the conditions
\trv\ imply that {\it the points where the entropy is maximized correspond
to points on the moduli space which are maximal intersection points of the marginal
stability walls for BPS states}.
The solutions to \trv\ can be characterized in purely geometric terms.
Let us consider the type IIB setup and look at the periods \aper\ - \bper\ of the Calabi-Yau.
Since we used the gauge $X^0 = 1$, the condition \triv\ means
that exactly $2h^{2,1}+1$ periods,
namely $(X^0, X^i; F_i)$ are real.
However, the last period $F_0 = {2 \over X^0} \CF_0  - {X^i\over X^0} F_i$
does not have to be real, since the phase of the prepotential is not fixed by
the phase of its derivatives.
In fact, if all of the periods were real, the holomorphic volume of the
Calabi-Yau $  \int \Omega
\wedge \bar \Omega= \big( X^I \bar F_I - \bar X^I F_I \big)$ would be zero.
Considering such  singular Calabi-Yau would hardly make sense, as it implies
that $\Omega$ is pointwise zero on the Calabi-Yau.
Fortunately, this is not the case since $F_0$ is not real.
Thus the geometry of periods is as given in Figure 1.

\ifig\periods{Calabi-Yau periods at the maximum entropy point
on the moduli space. }
{\epsfxsize3.2in\epsfbox{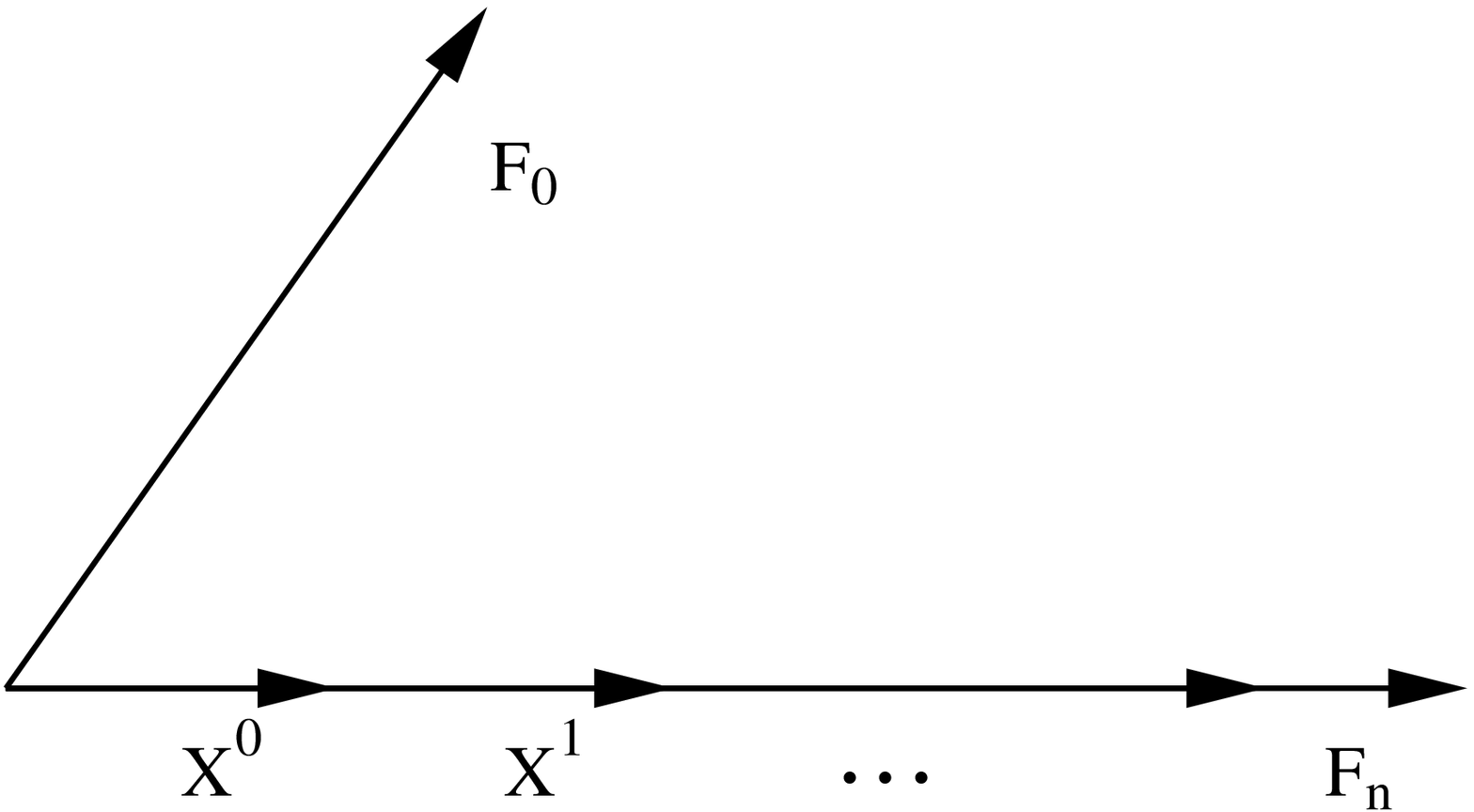}}

Note that a point where all periods but one are aligned is
a point where a maximal number of walls of marginal stability
for BPS states meet together.
In fact, this is the strongest condition we can have;
generically, it is impossible to have all periods aligned as the number
of constraints would be one higher than the number of parameters.
So, the condition for a maximum for the probability is the same
as maximal marginality for BPS bound states.

One can actually relax the condition $X^0=1$, and introduce
an arbitrary phase $X^0=e^{i\phi}$ instead. This will accordingly
rotate all other periods, resulting in an equivalent Calabi-Yau manifold.
Therefore, we can formulate an alternative
maximum criterion: {\it The entropy functional is maximized for the points on the
moduli space where all but one of the Calabi-Yau periods
are aligned on the complex plane, and $\i \tau_{ij} >0$.}

Suppose now that we can find such a point, where exactly $(2 h^{2,1}+1)$
of the periods are aligned.
We should stress that for a given prepotential these aligned
periods can actually be
some linear combinations of the canonical $A$ and  $B$ periods \aper -\bper.
Is there a freedom to choose, which of them we should use to fix
the normalization of $\Omega$ in the maximization problem \prob,
or there is a {\it canonical} choice of the cycle $A_0$?
The answer to the last question  is positive: the cycle $A_0$ is
dual to the 3-cycle which is not aligned with the rest of the 3-cycles.
In other words, $A_0$ corresponds to the null vector in the space
of $(2 h^{2,1}+1)$ aligned cycles with respect to the intersection pairing.
Thus, given the point on the moduli space where all but one
of the Calabi-Yau periods are aligned, the cycle $A_0$ is determined uniquely.
In the next section, we will illustrate this with a simple example of
the quintic three-fold.

Similarly, in the type IIA setup, we can consider bound states
of D0, D2, and D4 branes with charges $n_0$, $\vec n_2$, and $\vec n_4$, respectively.
The BPS mass of such states is given by the standard formula
\eqn\mbps{
M_{BPS} = |Z| = |n_0 + \vec n_2 \cdot \vec a
+ \vec n_4 \cdot \vec a^D|
}
Since in our case $\i a_i = \i a^D_i = 0$,
it follows from the BPS formula \mbps\ that
the critical points of type I are precisely
the points in the moduli space, where all bound states
of D0, D2, and D4 branes become marginal,
\eqn\bpscrit{
M_{BPS}(n_0, \vec n_2, \vec n_4)
= M_{BPS}(n_0,0,0) + M_{BPS}(0,\vec n_2,0) + M_{BPS}(0,0,\vec n_4)
}
%

\newsec{Examples}

In this section we will discuss two types of examples,
corresponding to   non-compact and compact  Calabi-Yau cases.
There is a crucial difference between these two cases in the type IIA setup.
Namely, on a compact Calabi-Yau manifold $M$, the cycles undergo
monodromies as one goes around singularities in the moduli space, while on
a non-compact Calabi-Yau there always exists at least one cycle
(0-cycle in the type IIA frame) which does not undergo
monodromy. Therefore, this is a canonical cycle to fix the period.
However, as we will see, the general approach based on \prob\
works in both cases.

As we discussed in the previous section, the problem
of finding  the maximum points on the moduli space
 is equivalent to the problem
of finding the points where all but one of the Calabi-Yau periods
are aligned. Unfortunately, at the moment it is unknown how to
find all such points for a given class of Calabi-Yau manifolds.
Our approach below is to look at the well known special points on
the moduli space (singularities, large complex structure, etc.) as
potential candidates. Therefore, our list of examples is far from
complete and serves just as an illustration of the general idea.
We find two types of solutions:  Solutions which correspond to
points on Calabi-Yau moduli which admit a structure similar to
complex multiplication.  The other type of solutions corresponds
to points on the moduli space where we have massless particles. In
the non-compact case the only set of examples where we actually
find a {\it maximum}, as opposed to minimum or other extremum
points, is when we have extra massless fields which lead to an
asymptotically free gauge theory.

\subsec{The Local $\cp^1$}

Let us start with the two simplest local models for a non-compact Calabi-Yau manifolds,
the total space of $\CO(-1)\oplus \CO(-1) \to \cp^1$ (conifold singularity), and
the total space of $\CO(0)\oplus  \CO(-2) \to \cp^1$.
(The exact stringy wave function for the conifold model
with an infinite set of non-normalizable deformations
was studied in \GSV.)
It is instructive to look first at the infinite product representation
of the topological string partition function on the A-model side \GoV.
For the conifold we have\foot{Here we use  topological string  conventions: $t = 2\pi(J-iB)$.
Sometimes an alternative convention $t=B + i J$ is used in the literature, since
then the mirror map is given by $t={X^1 \over X^0}$. In notations
 \ai\ then we have $a=B + i J$. }:
\eqn\infpmin{\eqalign{
Z_{(-1,-1)} = {\prod\limits_{n=1}^{\infty} (1-q^n Q)^n \over
\prod\limits_{n=1}^{\infty} (1-q^n )^n}
}}
where $q=e^{-g_s}$ and $Q = e^{-t}$, while for
$\CO(0)\oplus  \CO(-2) \to \cp^1$ case:
\eqn\infpmax{
Z_{(0,-2)} = {\prod\limits_{n=1}^{\infty} (1-q^n )^n \over
\prod\limits_{n=1}^{\infty} (1-q^n Q)^{n}}
}
In the semiclassical limit $q \to 1$, which is the
limit we are most interested in, the above expressions depend on $Q$ only.  Moreover,
 the first expression decreases while the second increases rapidly as $Q \to 1$.
 And as we will see in a moment,
the entropy functional has a minimum  for the conifold, and
a maximum for $\CO(0)\oplus  \CO(-2) \to \cp^1$ at $t=0$. Indeed,
since $Z = \exp \big( {{1\over g^2_s}F_{top}}\big)$, the entropy functional \entropya\ in
this representation is given by
%
\eqn\entrinf{\eqalign{
e^S = \big| Z\big|^2 e^{ - {1\over 2 }(t +\bar t)\big(
{\p \over \p t } \log Z + {\p \over \p \bar t } \log \bar Z \big)}
}}
As expected,  the second variation of the functional at the extremum is equal to
\eqn\svarinf{
{\d^2 S\over \d t \d\bar t } = -{2\pi\over g_s^2} \i \tau
}
where
\eqn\ztau{
\tau =i {g_s^2\over 2\pi}  {\p^2 \over \p t^2 } \log Z
}
At the conifold point we have $t =t^D =0$. Therefore \triv\  is satisfied and the conifold point
is an extremum. Moreover, near  $t=0$  we have
\eqn\conifoldpt{\eqalign{
 \tau = {i \over 2\pi } \log t + \ldots
}}
Hence, $\i \tau < 0$, which means that the conifold
point is a minimum.
On the contrary, since the infinite product expression \infpmax\
for the total space of  $\CO(0)\oplus  \CO(-2) \to \cp^1$ is given by the inverse
of \infpmin, in this case we have $\i \tau > 0$ at $t=0$.
Therefore, this is a maximum point.

One can arrive to the same conclusion by looking
at the genus-zero prepotential, which is equivalent
to the approach based on the infinite-product formula \infpmin.
For example, for the conifold, we have
\eqn\prconf{
F_{top}^0 = {1\over 12}t^3 - {\pi^2\over 6} t + \zeta(3)-\sum_{n=0}^{\infty}
{e^{-nt} \over n^3}
}
and therefore

\eqn\tauconif{ \tau = {i\over 2 \pi} \log (1 - e^{-t}) + \ldots
}
in agreement with \conifoldpt. For the total space of
$\CO(0)\oplus  \CO(-2) \to \cp^1$ we get the same expression with
an extra minus sign.

The difference between these two examples is very instructive:
At the conifold point, where the entropy in minimized, we  have extra massless
hypermultiplet,
while  at $t=0$ locus of $\CO(0)\oplus  \CO(-2) \to \cp^1$,
where the entropy in maximized,
an extra massless vector multiplet appears. The morale of the story
is that Calabi-Yau manifold providing the vector multiplet is preferred.

Let us conclude with a few general remarks.
Suppose we use topological strings to obtain
information about the effective supersymmetric four-dimensional
gauge theories with matter fields
and interactions which, in principle,
can have applications to phenomenological models of particle physics.
This can be done, for example, in the context of
the geometrical engineering program in string theory.
Here, topological strings provide a nice laboratory,
since many exact results for the topological string partition
function are available.

In the example we just studied, the distinction between these two
cases directly corresponds to whether or not the effective field
theory is asymptotically free. In particular, the appearance of a
vector multiplet, which corresponds to an asymptotically free
theory, is preferred over the non-asymptotically free theory, where
we have an extra massless hypermultiplet. From the effective field
theory viewpoint, this is translated into the statement that {\it
asymptotically free effective theories are preferred}. One can also
study other special points in moduli space, such as points which
correspond to conformal fixed points. In general, these will involve
more moduli parameters, and one might expect that in these cases we
will have a mixture of the above results:  along some directions the
probability for conformal fixed point is maximized and along others
it is minimized. In other words, it is an extremum for the entropy
functional, but not a definite maximum or minimum.  In fact we will discuss
an explicit example which corresponds to a multi-parameter model
where one finds that it is an extremum but not
a pure maximum or a minimum (see sec. 4.4 below).

\subsec{Large Complex Structure Limit}

Below we will argue that near the large complex structure limit in
the one-parameter Calabi-Yau models there is an infinite family of
solutions to the extremum equations \cm, labeled by an integer
number. Namely, for the values of the complexified K\"ahler
parameter $t$, such that $\r t =0$ and $ \i t \gg 1$, there are
infinitely many points where exactly three of the periods are
aligned. In a sense, these are the points where the Calabi-Yau
admits a deformed complex multiplication. Indeed, at these points,
to the leading approximation, $\tau$ satisfies a quadratic
equation with integer coefficients. The entropy functional
\entropy\ has a maximum for all such points.

Let us start with a simple model with the cubic prepotential
\eqn\lcsl{
\CF_0 = -{1\over 3} {(X^1)^3\over X^0}
}
This model captures the leading
behavior of all one-parameter Calabi-Yau models in the large complex structure limit.
It can also be viewed as the exact prepotential for some parts of the CY moduli
which receive no quantum corrections (such as the K\"{a}hler moduli of $T^2$'s inside CY threefolds).
The  vector of periods is given by
\eqn\plscl{\eqalign{
\pmatrix{ X^{0} \cr X^{1} \cr F_{1} \cr F_{0} }=
\pmatrix{  X^{0} \cr X^{1}  \cr -{(X^1)^2 / X^0} \cr
{1\over 3} {(X^1)^3 / (X^0)^2} }
}}
Let us consider these periods on the subspace $\i X^0 = 0$ and $\r X^1 =0$.
In other words, let us set the $B$-field to zero\foot{We use conventions,
where $t = a = {X^1 \over X^0} = B+i J$.}.
Then, two of the periods (namely, $X^1$ and $F_0$) are purely imaginary,
and two ($X^0$ and $F_1$) are real.
However, at the points where
\eqn\nper{
\Big( {X^1\over X^0}\Big)^2=n
}
for some (negative) integer $n$,
the linear combination of the periods $F^1+ n X^0$ vanishes!
This means that at these points the following three periods are aligned:
$(X^1, F_0, F^1+ n X^0)$;
they take purely imaginary values, while $X_0$ is real.
According to our general principle,
these points are extremum points of the entropy functional.
The main question now is whether this  extremum is a minimum or a maximum.

In order to answer this question we need to identify the 3-cycle $A_0$,
to find the resulting prepotential,
and to calculate the imaginary part of the
coupling constant. We use $SL(4, \Z)$ transformation
to bring the periods into the following form
\eqn\pertrans{\eqalign{
\pmatrix{ X^{0} \cr X^{1} \cr F_{1}+ n X^0 \cr F_{0} + n X^1 }=
\pmatrix{ 1 & 0& 0& 0 \cr
          0 & 1& 0& 0 \cr
          n & 0& 1& 0 \cr
          0 & n& 0& 1}
\pmatrix{ X^{0} \cr X^{1} \cr F_{1} \cr F_{0} }
}}
As we discussed in section 3.3, the cycle $A_0$ corresponds to
the null direction in the space spanned by the aligned periods.
In the present case, it is easy to see that the
period corresponding to this 3-cycle is $F_{0} + n X^1$.
Therefore, normalizing the value of the period of $\Omega$ over
this 3-cycle to unity, as in  \prob, is
equivalent to rescaling of all the periods by $\big( F_{0} + n X^1 \big)^{-1}$.
Then, the period vector takes the form:
\eqn\pernew{
\pmatrix{  -2 \tilde \CF_0 + \tilde X^1 \tilde F_1 \cr
- \tilde F_1 \cr
\tilde X^1 \cr
1}
= \pmatrix{ {1\over nt+ t^3/3} \cr
{t \over nt+ t^3/3} \cr
{-t^2+n \over nt+ t^3/3} \cr
1 }
}
where $t= {X^1 \over X^0}$ and $\tilde \CF_0$ is the prepotential
in the new basis. In these variables, the coupling constant
 is given by $\tau = {\p \tilde F_1 \over \p \tilde X^1}$.
Straightforward calculation    gives:
\eqn\taun{
\tau =  {i\over 4 \sqrt{|n|} }
}
at the critical point \nper, where $t = i \sqrt{|n|}$. Thus, all the
solutions from the infinite family \nper\
have $\i \tau > 0$ and correspond to the maxima of the entropy functional.
This simple example illustrates the behavior of a one-parameter
Calabi-Yau near the large complex structure limit, $\i t \gg 1$.
In this limit, prepotential receives small instanton corrections and
contains subleading quadratic and linear terms. However, in principle
one could still find the infinite family of aligned periods
by solving appropriately  modified equation \nper:
\eqn\nprmod{
\r F_1 + n X^0 = 0,
}
since the periods $X^1$ and $F_0$ in general case are
imaginary along the zero $B$-field  line $\r X^1=0, \, \i X^0 = 0$.
These will correspond, in the large $\i t$ limit to small perturbations
of the solutions
we have already discussed above.

However, we do not get interesting effective field theories at
such points, and the relative weight ("probability") of such
points is much smaller than that of the maxima where new massless
degrees of freedom appear, because of the logarithmic behavior
\conifoldpt\ of the coupling constant at the singular points.

\subsec{The Quintic}

Let us consider, following \COGP, the type IIA superstrings on the well studied compact Calabi-Yau
manifold with $h^{2,1}=1$,
which is the mirror of the quintic hypersurface in $\cp^4$.
It can be obtained as a $(\Z_5)^3$ quotient of the special quintic
\eqn\qiunt{ \sum_{i=1}^{5} z^5_i - 5 \psi \prod_{i=1}^{5} z_i = 0
}
The complex moduli space is $\cp^1$, parameterized by $z=\psi^{-5}$,
with three special points:
\eqn\thrpts{\eqalign{
& z=0:\quad\quad{\rm large~complex~structure~limit} \cr
& z=1:\quad\quad{\rm
conifold~point} \cr
& z=\infty:\quad\quad{\rm Gepner~point}
}}
The four periods undergo monodromy about these three points.
It is convenient to use the basis of the periods corresponding
to the BPS state in the mirror A-model labeled by the $(D6, D4, D2, D0)$-brane charges:
$(\Pi_{D0}, \Pi_{D2},\Pi_{D4},\Pi_{D6})$. These periods provide
corresponding D-brane tensions.
The general expression for the prepotential is
\eqn\prplcs{ F(t) = -{5 \over 6} t^3 -{11\over 4} t^2 + {25\over 12} t-
{25 i \over 2 \pi^3} \zeta(3) +
i \sum_{k=1}^{\infty} {d_k \over (2 \pi k)^3} e^{2 \pi i k t }
}
where
\eqn\mm{
t= {\Pi_{D2} \over \Pi_{D0}} =
 {1\over 2\pi i }\log z + \ldots
}
is a mirror map,
 and $d_k$ are instanton amplitudes (Gromov-Witten invariants), related
to the number $n_k$ of  rational curves of degree $k$ embedded in the
quintic, as
\eqn\dknk{
\sum_{k=0}^{\infty}  d_k e^{2 \pi i k t } =5 +
\sum_{k=1}^{\infty} {n_k k^3 \over 1-e^{2 \pi i t }} e^{2 \pi i k t }
}
We are interested in the solutions to the extremum equations
\cm. As was discussed before, we expect
an infinite family of such solutions at zero $B$-field
in the large complex structure limit.
Indeed, the periods in the basis \aper-\bper\ are
given by
\eqn\pqlcs{\eqalign{
\pmatrix{ X^{0} \cr X^{1} \cr F_{1} \cr F_{0} } =
\pmatrix{
1 \cr
t \cr
-{5 \over 2} t^2 -{11\over 2} t + {25\over 12} -
 \sum\limits_{k=1}^{\infty} {d_k \over (2 \pi k)^2} e^{2 \pi i k t } \cr
{5 \over 6} t^3 + {25\over 12} t-
{25 i \over  \pi^3} \zeta(3) +
 \sum\limits_{k=1}^{\infty} (2 i + 2 \pi k t){d_k \over (2 \pi k)^3} e^{2 \pi i k t }}
}}
Therefore, at the special set of points, where  $\r t = 0$ and a deformed
CM-type equation holds:
\eqn\qinsol{\eqalign{
{5 \over 2} t^2  - {25\over 12} +
 \sum\limits_{k=1}^{\infty} {d_k \over (2 \pi k)^2} e^{2 \pi i k t } = n
}}
the following  three periods are aligned:
$(X^1, F_0, F^1+ n X^0)$, where $n$ is an integer. It is clear that when $n \gg 1$ the
instanton corrections in \qinsol\ are small and general behavior
is similar to the cubic prepotential case. Therefore,
we conclude that this infinite set of solutions describes
local maxima of the entropy functional.

As we discussed earlier, the conifold point $z=1$ is a potential
candidate for a critical point of the entropy functional.
Let us then look at the periods
and try to find out which of them are  aligned.
The  periods  satisfy corresponding
Picard-Fuchs differential equation of hypergeometric type:
\eqn\qpf{
\Big[ \th_z^4 - \big(\th_z+{1\over 5}\big) \big(\th_z+{2\over 5}\big)
\big(\th_z+{3\over 5}\big) \big(\th_z+{4\over 5}\big)\Big] \Pi_i = 0
}
where $\th_z = z{ d\over dz}$. We use the conventions of \dgr\
to write down the basis of the solutions
to \qpf\ as follows:
\eqn\pbas{\eqalign{
&\Pi_0(z)=U_0(z) \cr
&\Pi_1(z)= U_1(z) \ {\rm if} \ \i z<0, \quad  U_1(z) + U_0(z) \ {\rm if} \ \i \ z>0 \cr
&\Pi_2(z)=U_2(z) \cr
&\Pi_3(z)= U_3(z) \ {\rm if} \ \i z<0, \quad  U_3(z) + U_2(z) \ {\rm if} \ \i \ z>0
}}
where $U_i$ are given in terms of the Meijer G-function \bateman
\eqn\pbas{\eqalign{
& U_0(z) = c \ G^{1,4}_{0,3}
\left( - z \left|\matrix{&  {4 \over 5} & {3 \over 5} & {2 \over 5} & {1 \over 5}\
\cr & 0 & 0 & 0 & 0 \cr} \right. \right)  \cr
& U_1(z) = {c \over 2\pi i} G^{2,4}_{1,2}
\left( z \left|\matrix{&  {4 \over 5} & {3 \over 5} & {2 \over 5} & {1 \over 5}\
\cr & 0 & 0 & 0 & 0 \cr} \right. \right)   \cr
& U_2(z) = {c \over (2\pi i)^2} G^{3,4}_{1,1}
\left( - z \left|\matrix{&  {4 \over 5} & {3 \over 5} & {2 \over 5} & {1 \over 5}\
\cr & 0 & 0 & 0 & 0 \cr} \right. \right)   \cr
& U_3(z) = {c \over (2\pi i)^3} G^{4,4}_{1,0}
\left( z \left|\matrix{&  {4 \over 5} & {3 \over 5} & {2 \over 5} & {1 \over 5}\
\cr & 0 & 0 & 0 & 0 \cr} \right. \right)
}}
and
\eqn\cdef{
c={1\over \Gamma \big( {1\over 5 }\big) \Gamma \big( {2\over 5 }\big)
\Gamma \big( {3\over 5 }\big) \Gamma \big( {4\over 5 }\big)}
}
Near $z=0$ the periods $\Pi_j$ behave as $(\log z)^j$. One can go to
another natural basis,
corresponding to $(D6, D4, D2, D0)$-brane state with the help of the
following transformation matrix
\eqn\ab{\eqalign{
\pmatrix{ \Pi_{D6} \cr \Pi_{D4} \cr \Pi_{D2} \cr \Pi_{D0} } =
\pmatrix{ 0 & 5& 0& 5 \cr 0 & 1& -5& 0 \cr 0& -1& 0& 0 \cr 1& 0& 0& 0}
\pmatrix{ \Pi_{0} \cr \Pi_{1} \cr \Pi_{2} \cr \Pi_{3} }
\quad\quad\quad\quad
}}
The intersection form in the  new basis is defined by $\#(D6 \cap D0)=1$ and
 $\#(D2 \cap D4)=1$.

Straightforward calculation at the conifold point\foot{We fix this
choice of the branch cut by requiring that D6 brane become massless
at the conifold point.}  $z=e^{-i0}$
with the help of the {\it Mathematica} package gives:
\eqn\percnf{\eqalign{
\pmatrix{ \Pi_{D6} \cr \Pi_{D4} \cr \Pi_{D2} \cr \Pi_{D0} } =
\pmatrix{ 0 \cr 5\alpha -7 i \beta \cr 2i\beta  \cr \gamma }
}}
where $(\alpha, \beta, \gamma)$ are real constants:
\eqn\abg{\eqalign{
&\alpha = - {\sqrt 5 \over 16 \pi^4} \, \r \, G^{3,4}_{0,1}
\left( - 1 \left|\matrix{&  {1 \over 5} & {2 \over 5} & {3 \over 5} & {4 \over 5}\
\cr & 0 & 0 & 0 & 0 \cr} \right. \right)\approx -1.239 \cr
&\beta= {\sqrt 5 \over 16 \pi^3} \, G^{2,4}_{0,2}
\left( 1 \left|\matrix{&  {1 \over 5} & {2 \over 5} & {3 \over 5} & {4 \over 5}\
\cr & 0 & 0 & 0 & 0 \cr} \right. \right)\approx 0.646787 \cr
&\gamma={_4} F_3\big( {1 \over 5},{2 \over 5},{3 \over 5},{4 \over 5};1,1,1;1  \big)
\approx 1.07073
}}
{}From \percnf\ we see that it is possible to align  three periods
by choosing  appropriate linear  transformation.
For example, we can take
\eqn\trnsf{\eqalign{
\pmatrix{ 2 & 0& 0& 0 \cr 0 & 2 & 7 & 0 \cr 0& 0& 1& 0 \cr 0& 0& 0& 1}
\pmatrix{ \Pi_{D6} \cr \Pi_{D4} \cr \Pi_{D2} \cr \Pi_{D0} } =
 \pmatrix{ 2 \Pi_{D6} \cr 2 \Pi_{D4} + 7 \Pi_{D2} \cr \Pi_{D2} \cr \Pi_{D0} } =
\pmatrix{ 0 \cr 10 \gamma  \cr 2 i\beta  \cr {\alpha }}
}}
Therefore, the conifold point in the quintic is a solution
to the extremum equations \cm\ if we fix
appropriate 3-cycle.
In particular, the choice \trnsf\ corresponds to fixing $X^0 = 2 \Pi_{D4} + 7 \Pi_{D2}$.

\subsec{A Multi-parameter Model}

Finally, we consider an example of a non-compact Calabi-Yau manifold
with several moduli fields. Such models exhibit some new phenomena.
For example, there can be points in the moduli space
where several different periods vanish,
and the corresponding BPS states become massless.
In general, one might expect such points to be saddle points
for the entropy functional (neither maxima nor minima). This is
indeed what we find in a specific example considered below.

\ifig\toricgraph{The toric diagram of the 3-parameter model.}
{\epsfxsize2.5in\epsfbox{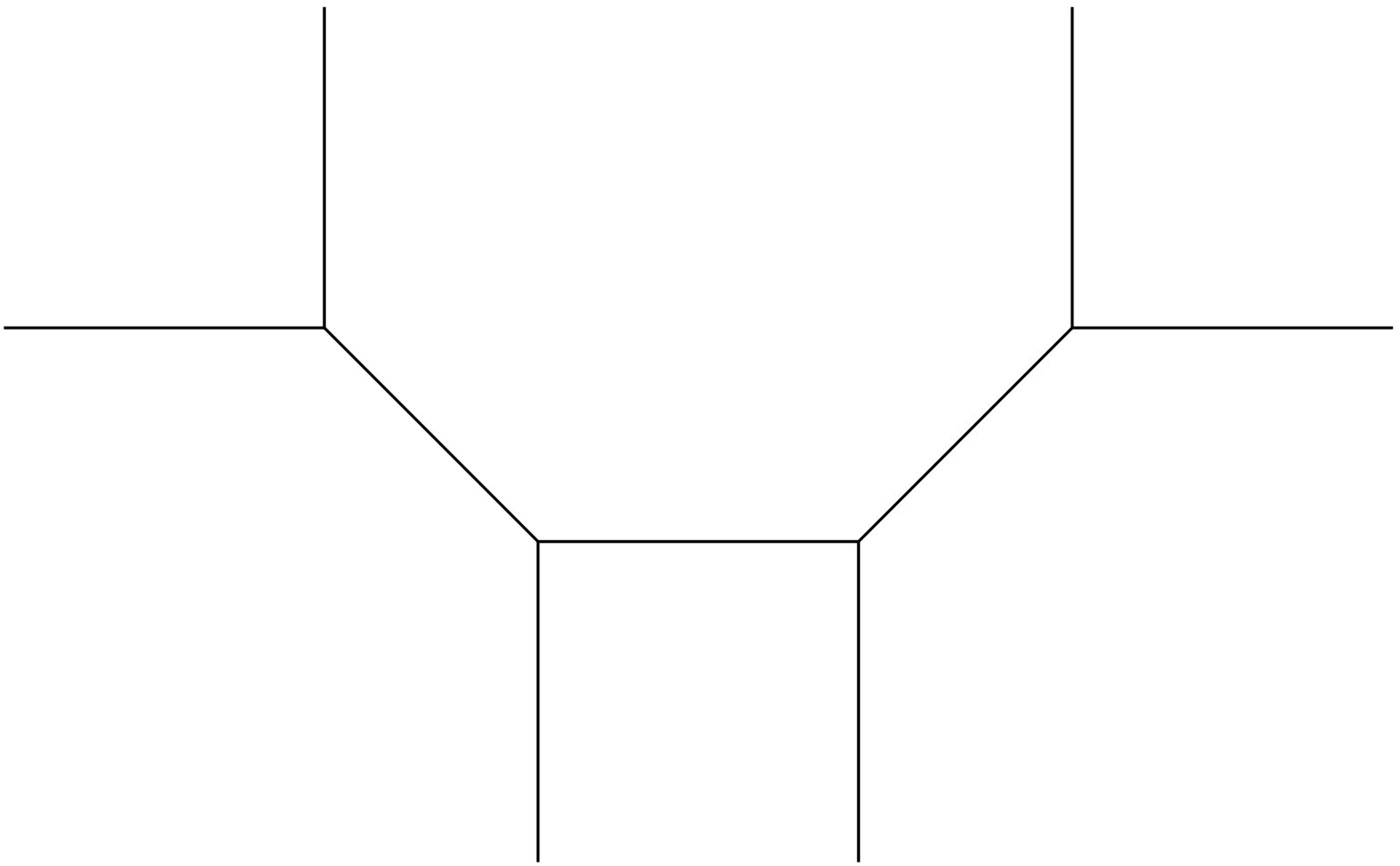}}

Consider a 3-parameter model
studied {\it e.g.} in \AMV. It has three 2-cycles, whose K\"{a}hler
parameters we denote\foot{In notations \ai, they are given by
$t_1 = - 2 \pi  i a^1, t_2 = - 2 \pi  i a^2, r = - 2 \pi  i a^3 $.} by $t_1$, $t_2$, and $r$.
The prepotential has the form:
\eqn\ttrprep{
F^0_{top} =  \sum_n {e^{-n t_1} \over n^3}  \sum_n {e^{-n t_2} \over n^3}
+ \sum_n {e^{-nr} (1-e^{-n t_1}) (1-e^{-n t_2}) \over n^3} }
and the dual variables are
\eqn\tdtdtd{\eqalign{
&  t_1^D =  {1\over 4\pi^2} \sum_n {e^{-n t_1} (1- e^{-nr} (1-e^{-n t_2})) \over n^2} \cr
& t_2^D =   {1\over 4\pi^2} \sum_n {e^{-n t_2} (1- e^{-nr}  (1-e^{-n t_1})) \over n^2} \cr
& r^D=  {1\over 4\pi^2} \sum_n {e^{-nr} (1-e^{-n t_1}) (1-e^{-n t_2}) \over n^2}
}}
The coupling constant matrix $\tau_{ij}$ is given by \ztau\ and
has the following entries (symmetric in $i,j = 1,2,3$):
\eqn\tauijentries{\eqalign{
& \tau_{11} =- {i \over 2\pi } \log {(1-e^{-t_1}) (1-e^{-(r + t_1 + t_2)})
\over (1-e^{-(r + t_1)}) } \cr
& \tau_{22} =- {i \over 2\pi } \log {(1-e^{-t_2}) (1-e^{-(r + t_1 + t_2)})
\over (1-e^{-(r + t_2)}) } \cr
& \tau_{12} = -{i \over 2\pi } \log (1-e^{-(r + t_1 + t_2)}) \cr
& \tau_{13} =- {i \over 2\pi } \log {(1-e^{-(r + t_1 + t_2)})
\over (1-e^{-(r + t_1)}) } \cr
& \tau_{23} = -{i \over 2\pi } \log {(1-e^{-(r + t_1 + t_2)})
\over (1-e^{-(r + t_2)}) } \cr
& \tau_{33} = -{i \over 2\pi } \log {(1-e^{-r}) (1-e^{-(r + t_1 + t_2)})
\over  (1-e^{-(r + t_1)}) (1-e^{-(r + t_2)}) }
}}
Consider taking the limit $(t_1,t_2,r) \to 0$ along the imaginary line,
such that the ratios ${t_1 /r}$ and ${t_2 / r}$ are kept fixed.
Then imaginary parts of the dual variables \tdtdtd\ are zero. Therefore, it is a
particular solution to the extremum equations \triv.

In order to determine the behavior of the entropy functional near  this extremum,
we should  diagonalize imaginary part of the matrix $\tau_{ij}$ and look at the eigenvalues.
It is easy to see that in this limit it is given by:
\eqn\tled{\eqalign{
\i \tau = - {\log|\, r|  \over 2\pi }
\pmatrix{
1  + \CO(x)& 1+ \CO(x)&  \CO(x)   \cr
1 + \CO(x) & 1+ \CO(x)& \CO(x) \cr
\CO(x)  & \CO(x)&  - 1 + \CO(x) }
}}
where $x \sim  \log ^{-1}|\, r| $.
To solve the diagonalization problem to the leading order in $x$, it is enough to
consider the matrix of the form
\eqn\mab{\eqalign{
\pmatrix{
1 & 1&  a   \cr
1  & 1& b \cr
a & b &  - 1  }
}}
where $a \ll 1$ and $b \sim a$.
The eigenvalues of this matrix  are given by $(1,{1\over 2}(a-b)^2,-2)$ to the leading order.
Therefore, the imaginary part of the coupling matrix \tled\ near the extremum point
has one large positive eigenvalue of order $\log|\, r|$,  one positive eigenvalue of order $1$, and
one large negative eigenvalue
of order $\log|\, r|$.
 Notice that  ${\rm sign} (\i \tau) =(2,1)$ and therefore, this is
an example of a signature of type $(h^{2,1}-1,1)$. In this case
we are having a saddle point of the entropy functional.

\newsec{Conclusions and Further Issues}

In this paper we discussed the behavior of the stringy wave function
on the moduli space of a Calabi-Yau manifold. It became a meaningful
quantity once we fixed a particular combination of charge/chemical potential
for one of the magnetic/electric charges of the black hole.
The square of this wave function can be interpreted as a measure
for string compactifications.
As we discussed, the solution to finding maxima/minima of this function has
a nice geometric meaning: they correspond to points on the moduli
space where all but one period of the holomorphic 3-form $\Omega$
have equal phase. The formulation of this geometric problem involves
a choice of a 3-cycle $A_0 \in H_3 (M,\Z)$, whose period we denote $X^0$
(or, a choice of $A_0 \in H^{{\rm even}} (M,\Z)$ in type IIA theory).

While it appears to be a rather challenging problem to obtain a complete
solution to these equations, we managed to find a certain class of solutions.
They fall into two families:  They either correspond to `quantum deformed'
complex multiplication points on the moduli space of a Calabi-Yau manifold,
or to points with extra massless particles.  Moreover, for the examples
with extra massless degrees of freedom the maxima
that we found correspond to the
points where the effective
field theory is asymptotically free.

As discussed above, in order to write down our wave function,
we need to choose a particular direction in the charge lattice.
In the type IIB case, this corresponds to choosing an integral 3-cycle.  It
is natural to ask how our conclusions depend on this choice (for the type IIA on non-compact CY
there is a natural direction of the charge lattice which corresponds
to D0 brane charge).
For a compact manifold $M$ there is no natural choice
of $A_0 \in H_3 (M,\Z)$. In fact, even for a particular choice of $A_0$,
there is an ambiguity related to the monodromy action on $H_3 (M,\Z)$.
To classify these choices we need to study the monodromy group
action on $H_3 (M,\Z)$.
For a Calabi-Yau manifold $M$, the monodromy group $H$
is a subgroup of $G = Sp(2h^{2,1} + 2, \Z)$, so that
the number of distinct choices of a 3-cycle is given
by the index $[G:H]$.
The calculation of $[G:H]$ for a compact Calabi-Yau space
is an interesting and challenging problem. By analogy with
the mapping class group of a genus-$g$ Riemann surface \Riema,
we may expect that $[G:H]$ is finite. In fact, the monodromy
group $H$ can be generated by two elements, which correspond
to monodromies around the conifold point and infinity.
For example, for the quintic threefold, we have
$$
M_c = \pmatrix{ 1 & 0& 0& 0 \cr 0 & 1& 0& 0 \cr -1& 0& 1& 0 \cr 0& 0& 0& 1}
\quad\quad\quad\quad
M_{\infty} = \pmatrix{ 1& -1& 5& -3 \cr 0& 1& -8& -5 \cr 0& 0& 1& 0 \cr 0& 0& 1& 1}
$$
In this case, it is easy to see that $[G:H] > 1$.

Another important question that deserves further study
is classification of the extremum points on the moduli space,
as solutions to the equations \triv. In mathematical terms,
the problem is to find all the points on the moduli space
where all but one of the
Calabi-Yau periods are aligned.

Finally it would be interesting to
understand more physically what it means to fix a charge/chemical
potential, and why that is natural.  It is conceivable
that this becomes natural in the context of decoupling gravity
from gauge theory.  In particular a preferred direction may be
the direction corresponding to the graviphoton charge.   It is worthwhile
trying to dynamically explain such a formulation of the problem.


\vskip 30pt

\centerline{\bf Acknowledgments}

We would like to thank R. Dijkgraaf, S. Katz, A. Klemm, M.
Ro\v{c}ek, T. Oliker and H. Ooguri for valuable discussions. We
also thank B.Fiol for pointing out a sign error in the large
complex structure limit example, that appeared in the first
version of the paper.

This research was supported in part by NSF grants PHY-0244821
and DMS-0244464. This work was conducted during the period
S.G. served as a Clay Mathematics Institute Long-Term Prize Fellow.
K.S. and S.G. are also supported in part by RFBR grant 04-02-16880.
We would like to thank the 2005 Simons Workshop on Mathematics
and Physics for providing a stimulating environment where part of
this work was done.
S.G. would also like to thank the KITP at Santa Barbara for hospitality
during the completion of this work. While at KITP, the research of S.G.
was supported in part by by the NSF under grant PHY99-07949.

\listrefs
\end